\documentstyle[twoside,fleqn,npb,epsfig]{article}
\newcommand{\AmS}{{\protect\the\textfont2
  A\kern-.1667em\lower.5ex\hbox{M}\kern-.125emS}}
\newcommand{\SUNC}[1]{
\mbox{\parbox{3cm}{
\begin{picture}(3.5,1.4)
\thicklines
\put(0.5,0.7){\line(1,0){2}}
\put(1.5,0.7){\circle{1}}
\put(2.65,0.7){\makebox(0,0)[l]{$(#1)$}}
\end{picture}
}}
\hfill}

\newcommand{\GLASS}[1]{
\mbox{\parbox{3cm}{
\begin{picture}(3.5,1.4)
\thicklines
\put(0.5,0.7){\line(1,0){0.5}}
\put(3,0.7){\line(1,0){0.5}}
\put(1.5,0.7){\circle{1}}
\put(2.5,0.7){\circle{1}}
\put(3.65,0.7){\makebox(0,0)[l]{$(#1)$}}
\end{picture}
}}
\hfill}

\newcommand{\TRI}[1]{
\mbox{\parbox{3cm}{
\begin{picture}(3.5,1.4)
\thicklines
\put(0.5,0.7){\line(1,0){0.5}}
\put(1.5,1.2){\line(0,-1){1}}
\put(1.5,1.2){\line(1,0){1}}
\put(1.5,0.2){\line(1,0){1}}
\put(1.5,0.7){\circle{1}}
\put(2.65,0.7){\makebox(0,0)[l]{$(#1)$}}
\end{picture}
}}
\hfill}

\newcommand{\ABOX}[2]{
\mbox{\parbox{3cm}{
\begin{picture}(3.5,1.4)
\thicklines
\put(0.5,0.2){\line(1,0){2.4}}
\put(0.5,1.2){\line(1,0){2.4}}
\put(1,0.2){\line(0,1){1}}
\put(2,0.7){\circle{1}}
\put(3,0.7){\makebox(0,0)[l]{$(#1,#2)$}}
\end{picture}
}}
\hfill}

\newcommand{\CBOX}[2]{
\mbox{\parbox{3cm}{
\begin{picture}(3.5,1.4)
\thicklines
\put(0.5,0.2){\line(1,0){2.4}}
\put(1.2,0.2){\line(1,1){1}}
\put(0.5,1.2){\line(1,0){2.4}}
\put(1.2,0.2){\line(0,1){1}}
\put(2.2,0.2){\line(0,1){1}}
\put(3.05,0.7){\makebox(0,0)[l]{$(#1,#2)$}}
\end{picture}
}} 
\hfill}

\newcommand{\Pboxa}[2]{
\mbox{\parbox{3cm}{
\begin{picture}(3.5,1.4)
\thicklines
\put(0.5,0.2){\line(1,0){2.4}}
\put(1.7,0.2){\line(0,1){1}}
\put(0.5,1.2){\line(1,0){2.4}}
\put(1,0.2){\line(0,1){1}}
\put(2.4,0.2){\line(0,1){1}}
\put(3.05,0.7){\makebox(0,0)[l]{$(#1,#2)$}}
\end{picture}
}} 
\hfill}
\newcommand{\PBOXa}{
\mbox{\parbox{3cm}{
\begin{picture}(3.5,1.4)
\thicklines
\put(0.5,0.2){\line(1,0){2.4}}
\put(1.7,0.2){\line(0,1){1}}
\put(0.5,1.2){\line(1,0){2.4}}
\put(1,0.2){\line(0,1){1}}
\put(2.4,0.2){\line(0,1){1}}
\end{picture}
}} 
\hfill}
\newcommand{\Pboxb}[2]{
\mbox{\parbox{3cm}{
\begin{picture}(3.5,1.4)
\thicklines
\put(0.5,0.2){\line(1,0){2.4}}
\put(1.7,0.2){\line(0,1){1}}
\put(0.5,1.2){\line(1,0){2.4}}
\put(1,0.2){\line(0,1){1}}
\put(2.4,0.2){\line(0,1){1}}
\put(1.35,0.7){\circle{0.5}}
\put(1.35,0.7){\makebox(0,0)[c]{$1$}}
\put(3.05,0.7){\makebox(0,0)[l]{$(#1,#2)$}}
\end{picture}
}} 
\hfill}
\newcommand{\Pboxij}[2]{
\mbox{\parbox{3cm}{
\begin{picture}(3.5,1.4)
\thicklines
\put(0.5,0.2){\line(1,0){2.4}}
\put(1.7,0.2){\line(0,1){1}}
\put(0.5,1.2){\line(1,0){2.4}}
\put(1,0.2){\line(0,1){1}}
\put(2.4,0.2){\line(0,1){1}}
\put(1.35,0.7){\circle{0.5}}
\put(1.35,0.7){\makebox(0,0)[c]{$#1$}}
\put(2.05,0.7){\circle{0.5}}
\put(2.05,0.7){\makebox(0,0)[c]{$#2$}}
\end{picture}
}} 
\hfill}
\newcommand{\PBOXb}{
\mbox{\parbox{3cm}{
\begin{picture}(3.5,1.4)
\thicklines
\put(0.5,0.2){\line(1,0){2.4}}
\put(1.7,0.2){\line(0,1){1}}
\put(0.5,1.2){\line(1,0){2.4}}
\put(1,0.2){\line(0,1){1}}
\put(2.4,0.2){\line(0,1){1}}
\put(1.35,0.7){\circle{0.5}}
\put(1.35,0.7){\makebox(0,0)[c]{$1$}}
\end{picture}
}} 
\hfill}
\newcommand{\Pboxc}[2]{
\mbox{\parbox{3cm}{
\begin{picture}(3.5,1.4)
\thicklines
\put(0.5,0.2){\line(1,0){2.4}}
\put(1.7,0.2){\line(0,1){1}}
\put(0.5,1.2){\line(1,0){2.4}}
\put(1,0.2){\line(0,1){1}}
\put(2.4,0.2){\line(0,1){1}}
\put(1.7,0.7){\circle*{0.2}}
\put(3.05,0.7){\makebox(0,0)[l]{$(#1,#2)$}}
\end{picture}
}} 
\hfill}
\newcommand{\PBOXc}{
\mbox{\parbox{3cm}{
\begin{picture}(3.5,1.4)
\thicklines
\put(0.5,0.2){\line(1,0){2.4}}
\put(1.7,0.2){\line(0,1){1}}
\put(0.5,1.2){\line(1,0){2.4}}
\put(1,0.2){\line(0,1){1}}
\put(2.4,0.2){\line(0,1){1}}
\put(1.7,0.7){\circle*{0.2}}
\end{picture}
}} 
\hfill}

\newcommand{\Xboxa}[2]{
\mbox{\parbox{3cm}{
\begin{picture}(3.5,1.4)
\thicklines
\put(0.5,0.2){\line(1,0){2.4}}
\put(0.5,1.2){\line(1,0){2.4}}
\put(1,0.2){\line(0,1){1}}
\put(1.7,0.2){\line(1,1){1}}
\put(1.7,1.2){\line(1,-1){1}}
\put(3.05,0.7){\makebox(0,0)[l]{$(#1,#2)$}}
\end{picture}
}} 
\hfill}

\newcommand{\Xboxb}[2]{
\mbox{\parbox{3cm}{
\begin{picture}(3.5,1.4)
\thicklines
\put(0.5,0.2){\line(1,0){2.4}}
\put(1.7,0.2){\line(1,1){1}}
\put(0.5,1.2){\line(1,0){2.4}}
\put(1,0.2){\line(0,1){1}}
\put(1.7,1.2){\line(1,-1){1}}
\put(1,0.7){\circle*{0.2}}
\put(3.05,0.7){\makebox(0,0)[l]{$(#1,#2)$}}
\end{picture}
}} 
\hfill}

\newcommand{\Xtri}[1]{
\mbox{\parbox{3cm}{
\begin{picture}(3.5,1.4)
\thicklines
\put(0.5,0.7){\line(1,0){0.7}}
\put(1.2,0.7){\line(3,1){2.2}}
\put(1.2,0.7){\line(3,-1){2.2}}
\put(1.8,0.9){\line(3,-2){1.2}}
\put(1.8,0.5){\line(3,2){1.2}}
\put(3.05,0.7){\makebox(0,0)[l]{$(#1)$}}
\end{picture}
}} 
\hfill}

\title{Progress towards $2 \to 2$ scattering at two loops}

\author{E.W.N.~Glover,\address{Institute of Particle Physics Phenomenology
        Department of Physics,
        University of Durham, Durham DH1 3LE, England}\thanks{
        \tt E.W.N.Glover@durham.ac.uk}
        and
        M.E.~Tejeda-Yeomans{\hbox{$^{\rm a}$}}\thanks{
        \tt M.E.Tejeda-Yeomans@durham.ac.uk}
}

\begin{document}
\unitlength0.6cm

\begin{abstract}
We discuss the two-loop integrals necessary for evaluating 
massless $2 \to 2$ scattering amplitudes.  As a test process, we consider
the leading colour two-loop contribution to 
$q\bar q \to q^\prime\bar q^\prime$.  We show that for physical scattering
processes the two Smirnov-Veretin planar box graphs $I_1$ and $I_2$ 
are accompanied by 
factors of $1/(D-4)$ thereby
necessitating a knowledge of both $I_1$ and $I_2$ to ${\cal O}(\epsilon)$.
Using an alternative basis $I_1$ and the irreducible numerator
integral $I_3$, the factors of $1/(D-4)$ disappear.    
\end{abstract}

\maketitle

\thispagestyle{myheadings}
\markright{DTP/00/62, IPPP/00/03, hep-ph/0010031}

\section{Introduction}

Two-to-two scattering processes are well known to be   one of the most basic
probes of the fundamental interactions of nature.   In hadron-hadron
collisions, parton-parton scattering to form  a large transverse momentum jet
tests the pointlike nature of the partons down to distance scales of
$10^{-17}$~m.  However, extracting useful results from experimental data 
requires both plentiful data and accurate theoretical calculations.  For
example, the single jet inclusive transverse energy distribution observed by
the CDF collaboration in Run I at the TEVATRON indicated possible new physics
at large transverse energy \cite{CDF}.  Data obtained by  the D0 collaboration
\cite{D0} was more consistent with theoretical next-to-leading order
expectations, however, because of both  theoretical and experimental
uncertainties no definite conclusion could be drawn.  The experimental
situation may be clarified in the forthcoming high statistic Run II starting in
2001. The theoretical prediction may be improved by including the
next-to-next-to-leading order perturbative predictions.  This has the effect of
(a) reducing the renormalisation scale dependence and (b) improving the
matching of the parton level theoretical jet algorithm with the hadron level
experimental jet algorithm because the jet structure can be modelled by the
presence of a third parton. Varying the renormalisation scale up and down by a
factor of two about the jet transverse energy leads to a 20\% (10\%)
renormalisation scale uncertainty at leading order (next-to-leading order)
for jets with $E_T \sim 100$~GeV. The improvement in accuracy
expected at next-to-next-to-leading order can be estimated using the
renormalisation group equations together with the known  leading and
next-to-leading order coefficients and is at the 1-2\% level. 
Of course, the  full next-to-next-to-leading order prediction requires a
knowledge of the two-loop $2 \to 2$ matrix elements as well as the
contributions from the one-loop  $2 \to 3$ and tree-level $2 \to 4$
processes.   

In this talk, we wish to review the recent progress that has been made towards 
the analaytic evaluation of the two-loop matrix elements relevant for
massless $2 \to 2$ scattering.
As can be seen from Table~1, the number of Feynman diagrams
contributing to the basic parton scattering processes increases dramatically 
with the number of loops. The one-loop graphs are those computed by Ellis and
Sexton \cite{ES} in 1986.
\begin{table}[ht]
\begin{center}
\begin{tabular}{c| c| c | c }
 \hline
Process  &   Tree & One loop & Two loops \\
\hline
$gg\to gg$ & 4 & 72 & 1531 \\
$q \bar q \to gg$ & 3 & 29 & ~563  \\
$q\bar q \to q^\prime\bar q^\prime$ & 1 & 10 & ~186 \\
\hline  
\end{tabular}
\end{center}
Table~1. Numbers of Feynman diagrams contributing to $2 \to 2$ parton scattering processes
\end{table}
The much more numerous two-loop graphs may be either products of one-loop
graphs, self-energy insertions or genuinely new topologies. It is the latter
class
 which has proved 
to be a major stumbling block. However, in
the last twelve months, all of the necessary integrals have been computed and
a complete basis set of master integrals now exists.
We note in passing that a particular two-loop helicity amplitude for $gg \to gg$ scattering
has been calculated by Bern, Dixon and Kosower \cite{bdk}.

\section{Master Integrals}
The complete set of massless master integrals comprises the trivial topologies
of single scale integrals which can be written as products of Gamma functions, 
$$
\SUNC{s}  \hspace{-0.5cm} \GLASS{s} \hfill \TRI{s}
$$
the less trivial non-planar triangle graph \cite{xtri},
$$
\Xtri{s}
$$
two scale integrals that are related to the one-loop box
graphs \cite{AGO2},
$$
\ABOX{s}{t} \hfill \CBOX{s}{t} 
$$
the planar double boxes \cite{planar1,planar2}
$$
\Pboxa{s}{t} \hfill \Pboxc{s}{t} 
$$
which we denote $I_1$ and $I_2$ respectively 
together with the non-planar double boxes \cite{nonplanar1,nonplanar2},
$$
\Xboxa{s}{t} \hfill \Xboxb{s}{t} .
$$
The Mandelstam variables $s$ and $t$ represent the kinematic scales 
involved in the integral while
the blobs on the propagators represent an additional power of that propagator.
The latter blobbed graphs are necessary to evaluate tensor integrals.   In other words,
starting from a planar or non-planar box tensor integral, it is not possible to 
reduce the powers of all propagators to unity and the second master integral is
required \cite{planar2,nonplanar2}.
The scalar planar \cite{planar1} and non-planar \cite{nonplanar1} integrals 
themselves were evaluated as multiple Mellin-Barnes integrals and represent significant
achievements in the field of Feynman diagrammology.

\section{Application: $q \bar q \to q^\prime \bar q^\prime$}
As a precursor to 
a full two loop calculation of all massless $2 \to 2$ scattering matrix elements let us
concentrate on the specific process $q (p_1) \bar q (p_2)  \to \bar{q}^\prime (p_3)
 q^\prime (p_4)$ where the lightlike
momentum assignments are in parentheses. 

The amplitude ${\cal M}$ has the perturbative expansion,
$$
{\cal M}=g_s^2{\cal M}_0+g_s^4{\cal M}_1+g_s^6{\cal M}_2+ {\cal
O}(g_s^{8}),
$$
in terms of the tree level (${\cal M}_0$), one-loop (${\cal M}_1$) and 
two-loop (${\cal M}_2$) amplitudes.
The squared and summed tree-level amplitude is given by,
\begin{equation}
|{\cal M}_0|^2 = 2(N^2-1)\left (\frac{t^2+u^2}{s^2}-\epsilon \right)
\end{equation}
where $ s = (p_1+p_2)^2$, $t=(p_2-p_3)^2$ and 
$u=-s-t$.
The one-loop amplitude ${\cal M}_1$ 
was first calculated by Ellis and Sexton \cite{ES} and contributes to the cross
section at ${\cal O}(g_s^6)$, or next-to-leading order.
The two-loop amplitude ${\cal M}_2$ first contributes at
${\cal O}(g_s^8)$ through its interference with the
tree level amplitude ${\cal M}_0$ and has the following colour structure,
\begin{eqnarray*}
\lefteqn{{\cal M}_2 {\cal M}_0^\dagger + {\cal M}_2^\dagger {\cal M}_0
= \left (N^2-1\right) }\\
&\times&
 \left(
A N^2 + B + C \frac{1}{N^2} + D N_F N + E \frac{
N_F}{N} \right)
\end{eqnarray*}
where $N$ is the number of colours and $N_F$ the number of light fermions.
The leading colour amplitude $A$ is gauge invariant and
contains only planar diagrams such as those shown in Figure~1.
\begin{figure}[h]
   \begin{center}  
   \epsfysize=1.5cm
   \epsfxsize=2.5cm
   \epsffile{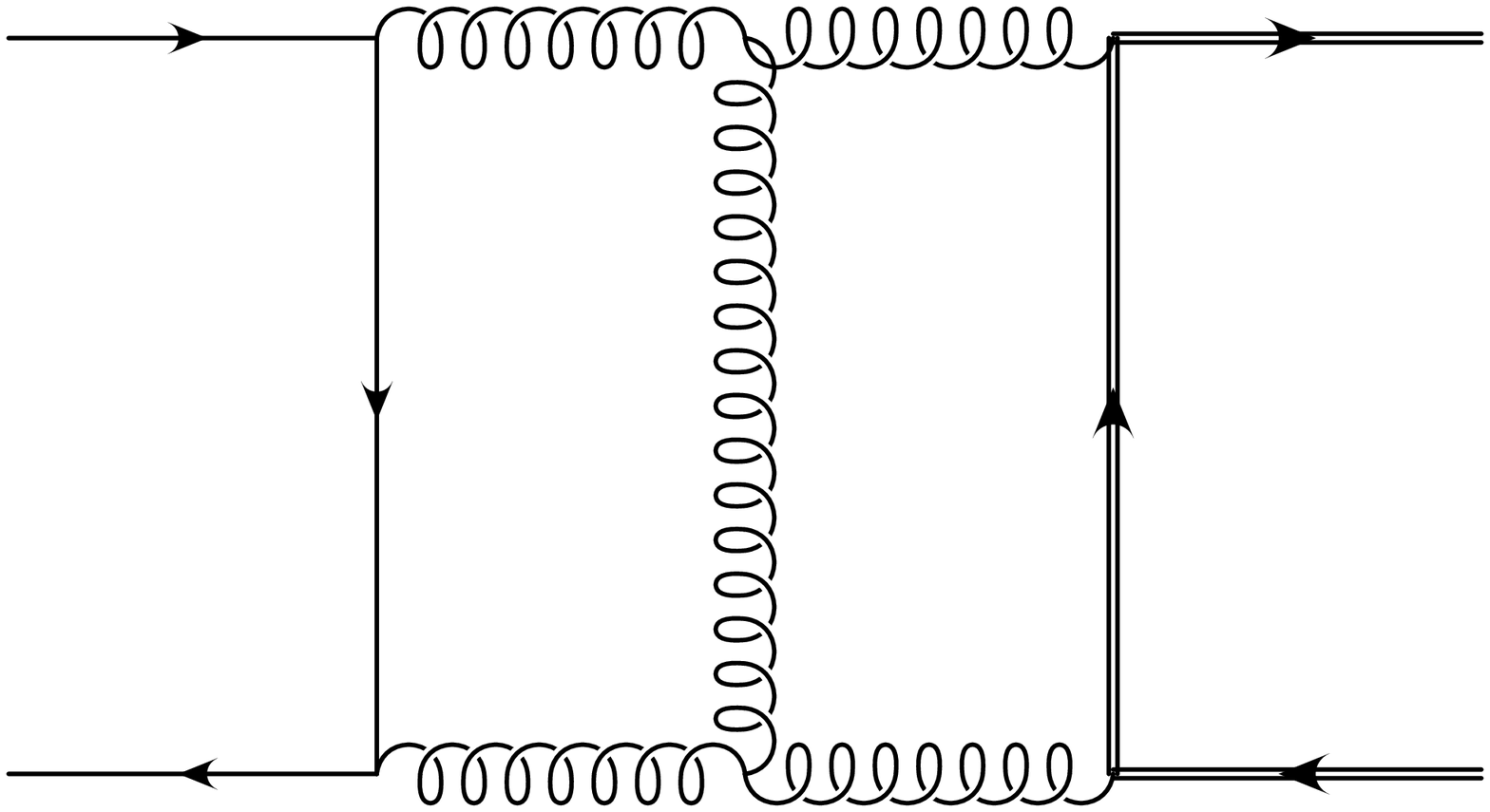}
   \hspace{1cm}
   \epsfysize=1.5cm
   \epsfxsize=2.5cm
   \epsffile{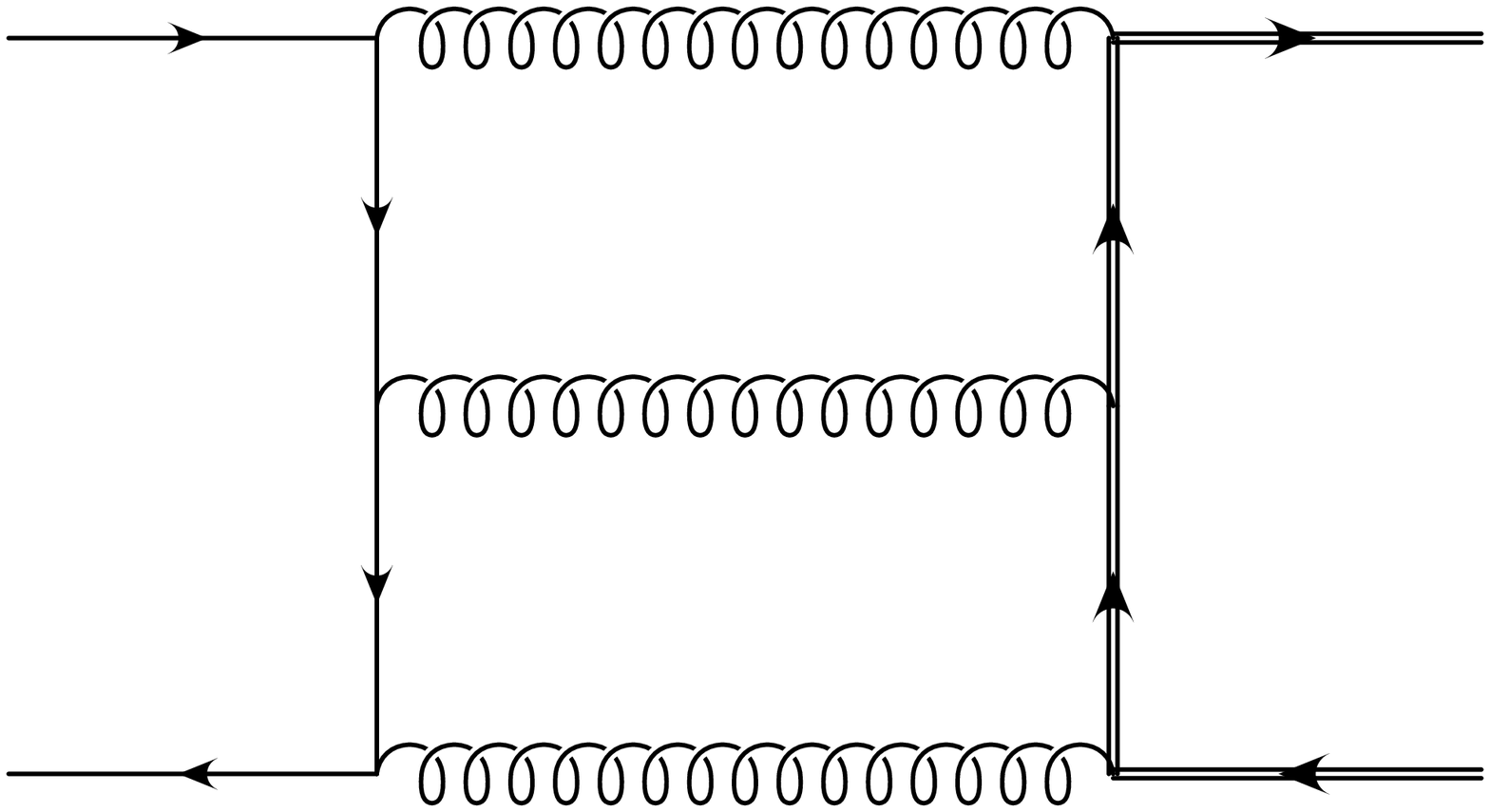}
\end{center}
Figure~1: The planar box graphs for $q \bar q \to q^\prime \bar q^\prime$
contributing at leading colour
\end{figure}
The amplitudes suppressed by powers of $N$ ($B, C, D$
and $E$) contain the
non-planar graphs.  As a first step in carrying out the analytic evaluation of
the two loop graphs, we therefore focus on the leading colour amplitude $A$. 

\section{Auxiliary diagram}
To handle all possible permutations of planar diagrams it is convenient to work
with the auxiliary diagram shown in Figure~2,
\vspace{-0.5cm}
\begin{figure}[h]
   \begin{center}  
   \epsfysize=4.0cm
   \epsfxsize=6.0cm
   \epsffile{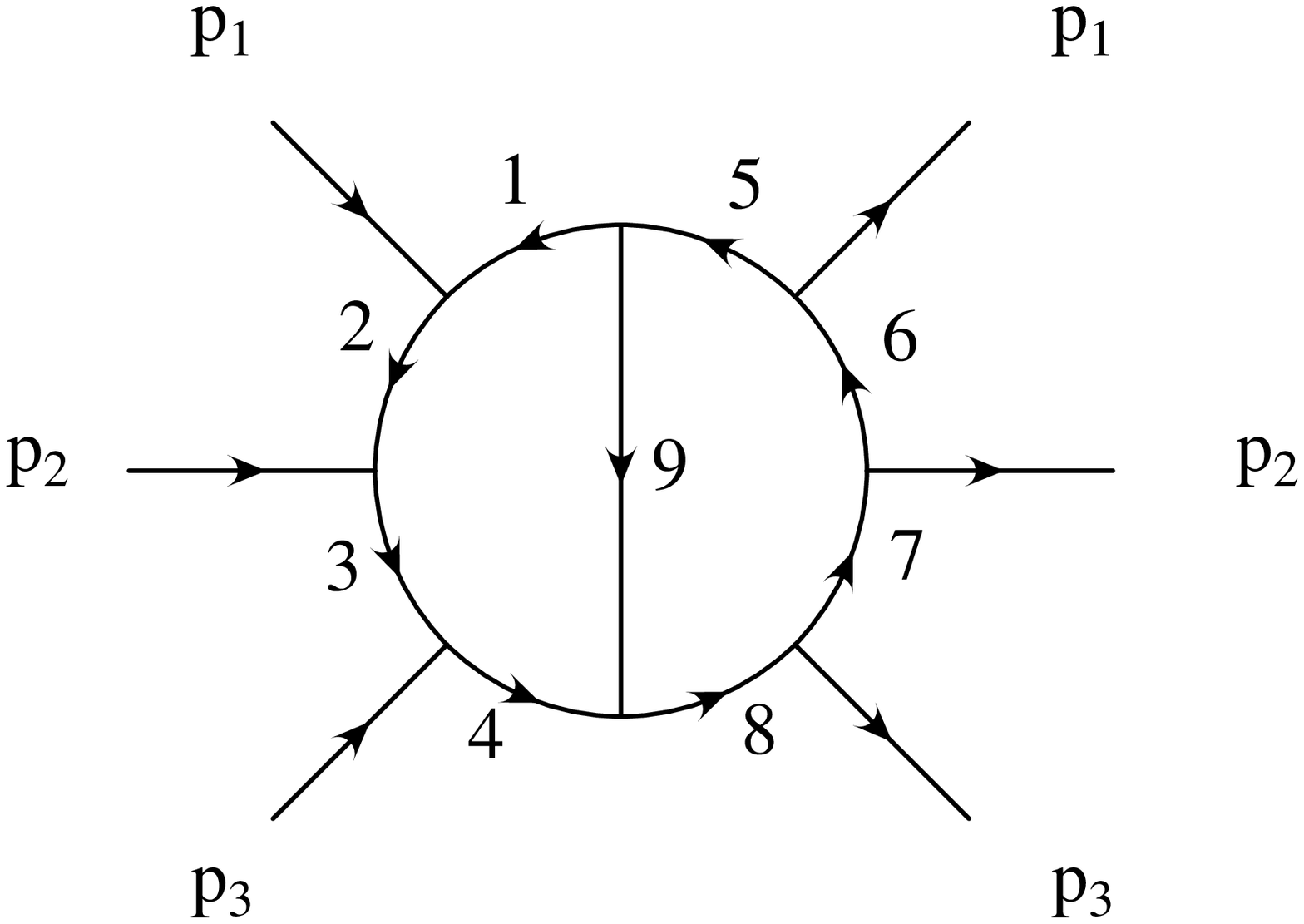}
\end{center}
Figure~2: The general auxiliary planar diagram with 
propagator $i$ raised to the power $\nu_i$
\end{figure}
\vspace{-0.5cm}
\begin{eqnarray*}
\lefteqn{I^D(\nu_1,\nu_2,\nu_3,\nu_4,\nu_5,\nu_6,\nu_7,\nu_8,\nu_9)=}\\
&& \hspace{-0.8cm}
\int 
\frac{d^Dk_1}{i\pi^{d/2}}
\int
\frac{d^Dk_2}{i\pi^{d/2}}
\frac{1}{A_1^{\nu_1}A_2^{\nu_2}A_3^{\nu_3}A_4^{\nu_4}A_5^{\nu_5}A_6^{\nu_6}
A_7^{\nu_7}A_8^{\nu_8}A_9^{\nu_9}}
\end{eqnarray*}
where $A_1 = k_1^2$, 
$A_2 = (k_1+p_1)^2$, 
$A_3 = (k_1+p_{12})^2$, 
$A_4 = (k_1+p_{123})^2$, 
$A_5 = k_2^2$,
$A_6 = (k_2+p_1)^2$,
$A_7 = (k_2+p_{12})^2$,
$A_8 = (k_2+p_{123})^2$, 
$A_9 = (k_2-k_1)^2$
and $p_{ij}=p_i+p_j$ and $p_{ijk}=p_i+p_j+p_k$. The $i$th propagator is raised to
the power $\nu_i$.  Scalar integrals have all $\nu_i = 1$ or 0.   For example,
\begin{eqnarray*}
\Pboxa{s}{t}  &=& I^D(1,1,1,0,1,0,1,1,1)\\
\CBOX{s}{t}  &=& I^D(1,1,0,0,0,0,1,1,1)\\
\ABOX{s}{t}  &=& I^D(1,1,1,0,0,0,0,1,1)
\end{eqnarray*}
By permuting the arguments, we obtain the other orientations. 
For example, 
$$
\Pboxa{t}{s}  = I^D(1,1,0,1,0,1,1,1,1).
$$
For the interference of tree-level with two-loop graphs,  loop momenta in the
numerator are always contracted with either external or loop momenta. These
dot-products can always be written as combinations of the 9 propagators so that
tensor integrals appear as generalised scalar integals. Negative values of
$\nu_i$ correspond to irreducible numerators.
For example, the planar box integral with one irreducible numerator on the left hand
loop can be written,
$$
\PBOXb  = I^D(1,1,1,-1,1,0,1,1,1).
$$

\section{General procedure}
The general procedure for computing the two loop graphs is as follows:
\begin{enumerate}
\item Use QGRAF \cite{QGRAF} to generate the Feynman diagrams
\item Multiply by tree-level and compute traces
\item Identify combinations of scalar and tensor auxiliary integrals
\item Exchange tensor integrals ($\nu_i < 0$) for auxiliary integrals
in higher dimension with higher powers of propagators ($\nu_i \geq 0$)
\cite{AGO3}
\item Apply integration-by-parts (IBP) identities \cite{IBP}
to reduce general auxiliary integrals to combinations of the Master Topologies 
$(\nu_i
\neq 1)$
\item Apply specific IBP identities to reduce Master Topologies to Master
Integrals $(\nu_i = 1)$ 
\end{enumerate}

\section{Planar box graphs}
However, because the specific IBP identities for the planar box are quite
complicated \cite{planar2}, we choose an alternative method for the  tensor
planar box graphs.  In particular we try to stay in $D\sim 4$. To do this we
adopt the approach of Gehrmann and Remiddi \cite{diffeq}.  The idea is very
simple both in concept and in implementation.   We characterise a loop integral
with three numbers, $t$ the number of different propagators in the denominator,
$r$ the sum of powers of propagators in the denominator and $s$ the sum of
powers of propagators in the numerator.    For example, 
\begin{eqnarray*}
\Pboxa{s}{t}  && t=r=7,~~s=0\\
\Pboxb{s}{t}  && t=r=7,~~s=1\\
\Pboxc{s}{t}  && t=7,~~r=8,~~s=0
\end{eqnarray*}

When acting on a loop integral $I_{t,r,s}$, the IBP (and Lorentz invariance \cite{diffeq}) 
identities produce more complicated integrals with the same topology
$I_{t,r+1,s}$ and $I_{t,r+1,s+1}$, 
simpler integrals with the same topology $I_{t,r-1,s}$ and $I_{t,r-1,s-1}$
as well as simpler topologies $I_{t-1,r,s}$.  By applying each identity to 
each $I_{t,r,s}$ , we can form a linear system of equations from which the more
complicated integrals can be eliminated. 

For example, when $t=7$, the number of integrals 
for a given value of $r$ and $s$ is shown in Table~2.
\begin{table}[h]
\begin{center}
\begin{tabular}{|c |l||r|r|r|r|r|}\hline
                 \multicolumn{2}{|c||} {}     &           \multicolumn{5}{|c|}{s}\\ \cline{3-7}
                  \multicolumn{2}{|c||}{}      &    0 &    1 &    2 &    3 &    4 \\ \hline\hline
       &   7   &    1 &    2 &    3 &    4 &    5 \\ \cline{2-7}
    r  &   8   &    7 &   14 &   21 &   28 &   35 \\ \cline{2-7}
       &   9   &   28 &   56 &   84 &  112 &  140 \\ \cline{2-7}
       &  10   &   84 &  168 &  252 &  336 &  420 \\ \hline 
\end{tabular}
\end{center}
Table~2: Numbers of integrals with different values of $r$ and $s$ for $t=7$
(taken from \cite{diffeq}).
\end{table}
The two master integrals \cite{planar1,planar2} have $t=7$ and $r=7$ and $r=8$
respectively.
Tensor integrals (corresponding to $r=7$ and $s>0$) lie on the first row of
Table~2.  Using only these integrals as seeds for the IBP identities and
eliminating the unknowns using linear algebra, we immediately obtain 
all tensor integrals for the planar box in $D\sim 4$ in terms of
$
I_1$ and the irreducible numerator graph
$$
 I_3 = \PBOXb 
$$
(rather than $I_1$ and $I_2$)
together with simpler
pinched integrals that can be straightforwardly simplified.
If we denote the $s=i+j$, $t=r=7$ planar box integral as
$$
 I^D(1,1,1,-i,1,-j,1,1,1) = \Pboxij{i}{j}
$$ 
then, 
$$\Pboxij{i}{j} \hspace{-1cm}= \Pboxij{j}{i},$$
and, for example, the second rank tensor integrals are given by,
\begin{eqnarray*}
\lefteqn{
\Pboxij{1}{1} \hspace{-1cm}= \frac{st}{2}\Pboxa{s}{t}}\\
&-&\frac{3s}{2}\Pboxb{s}{t}\\
&+&\frac{8(D-3)}{(D-4)}\ABOX{s}{t}\\
&-&\frac{(7s+9t)}{s}\CBOX{s}{t}\\
&+&\frac{17(D-3)(3D-10)}{2s(D-4)^2}\TRI{s}\\
&-&\frac{2(3D-8)(3D-10)}{s^2(D-4)^3}\SUNC{s}\\
&+&\frac{9(3D-10)(3D-8)(D-3)}{st(D-4)^3}\SUNC{t}\\
\lefteqn{\phantom{~}\hspace{-0.5cm}
+\frac{2(D-3)((2D-5)s+2(D-3)t)}{s^2(D-4)^2}\GLASS{s}}\\
~~~~\\
\lefteqn{
\Pboxij{2}{0} \hspace{-1cm}
=\frac{(D-4)st}{2(D-3)}\PBOXa}\\
&-&\frac{(3(D-4)s-2t)}{2(D-3)}\PBOXb\\
&+&8\ABOX{s}{t}\\
&-&\frac{(7s+9t)(D-4)}{s(D-3)}\CBOX{s}{t}\\
&+&\frac{(13s-2t)(3D-10)}{2s^2(D-4)}\TRI{s}\\
&+&\frac{2(s+2t)(3D-8)(3D-10)}{2s^3(D-3)(D-4)^2}\SUNC{s}\\
&+&\frac{9(3D-8)(3D-10)}{st(D-4)^2}\SUNC{t}\\
&+&\frac{2((2D-7)s+2(D-4)t)}{s^2(D-4)}\GLASS{s}
\end{eqnarray*}
We see that the coefficients of both $I_1$ and $I_3$ are always 
finite as $D \to 4$ and that the IBP identities have not introduced fake
singularities.

\section{Leading Colour Matrix Elements}

In $D$ dimensions, the leading colour two-loop amplitude $A$ for
$q \bar q \to q^\prime \bar q^\prime$ scattering has
the following structure,
\begin{eqnarray*}
\lefteqn{A=  -\frac{2}{(4\pi)^D} \Biggl (}\\
&&\hspace{-0.8cm}+ { a_1} \Pboxa{s}{t} + { a_2} \Pboxb{s}{t} \\
&&\hspace{-0.8cm}+ { a_3} \Pboxa{t}{s} + { a_4} \Pboxb{t}{s} \\
&&\hspace{-0.8cm}+ { a_5} \ABOX{s}{t} + { a_6} \ABOX{t}{s} \\
&&\hspace{-0.8cm}+ { a_7} \CBOX{s}{t} + { a_8} \CBOX{t}{s} \\
&&\hspace{-0.8cm}+ { a_{9}} \TRI{s} + { a_{10}} \TRI{t} \\
&&\hspace{-0.8cm}+ { a_{11}} \SUNC{s} + { a_{12}} \SUNC{t} \\
&&\hspace{-0.8cm}+ { a_{13}} \GLASS{s} + { a_{14}} \GLASS{t}
\hspace{-0.5cm}\Biggr ).
\end{eqnarray*}
We have calculated the coefficients $a_1$ --- $ a_{14}$ in arbitrary dimension.
For example, the first two coefficients of the planar box graphs are given by,
\begin{eqnarray*}
{ a_1} &=& -\frac{(11D^2-53D+54)}{4(D-3)} ~{ s^3t}
-  4 ~{ s t^3}\\
&&-\frac{(D^2+2D-18)}{(D-3)} ~{ s^2 t^2}\\
&&\\
{ a_2} &=& \frac{(15D^2-71D+69)}{2(D-3)} ~{ s^3}
+\frac{2(7D-24)}{(D-3)} ~{ s t^2}\\
&&+ \frac{4(D^2-12)}{(D-3)} ~{ s^2 t} \\
%&&\\
%{ a_3} &=& -\frac{(3D^3+16D^2-196D+368)}{64(D-3)} ~{ s^2 t^2}\\
%&&- \frac{(D-2)(9D-28)(3D^2-18D+32)}{128(D-3)^2} ~{ s t^3}
%-4 ~{ t^4}\\
%&&\\
%{ a_4} &=& \frac{(18D^4-165D^3+308D^2+708D-1904)}{64(D-3)^2} ~{ st^2}\\
%&&+ \frac{3(27D^4-300D^3+1028D^2-944D-512)}{128(D-3)^2} ~{ t^3}
\end{eqnarray*}
We see that both $a_1$ and $a_2$ are well behaved as $D \to 4$ indicating again
that the IBP identities have not introduced fake
singularities.

\section{Relationship between Master Integrals}
We can also use the Gehrmann-Remiddi approach \cite{diffeq}
to find a relation between the
irreducible numerator master integral $I_3$ and those of Smirnov and Veretin 
($I_1$ and $I_2$) \cite{planar1,planar2}.
Suppressing the simpler pinched integrals, we find,
\begin{eqnarray*}
\lefteqn{
\PBOXb\hspace{-1cm} 
= -\frac{(3D-14)s}{2{(D-4)}}\PBOXa}\\
&&-\frac{(D-6)st}{2{(D-4)}(D-5)}\PBOXc\\
&& + ~{\rm pinchings}.
\end{eqnarray*}
The presence of the $D-4$ factor in the denominator
immediately indicates a problem.   
Both $I_1$ and $I_2$ have
$1/\epsilon^4$ leading
poles and it would appear that $I_3 \sim 1/\epsilon^5$. This is
{\bf not} the case as close examination of $I_1$ and $I_2$ shows that in this
combination the $1/\epsilon^4$ poles and descendents cancel completely
so that
$$
\PBOXb\hspace{-1cm} \sim \frac{1}{\epsilon^4}
$$
as we expect.
However, the finite parts of $I_3$ 
are controlled by the ${\cal O}(\epsilon)$ parts of $I_1$ and
$I_2$.

$I_3$ can also be written in terms of derivatives of $I_1$,
\begin{eqnarray*}
\PBOXb \hspace{-1cm}&=& -\frac{(s(D-5)-t)}{(D-4)} \PBOXa \\
&&+ \frac{ut}{(D-4)} ~~\frac{\partial}{\partial t}\hspace{-0.3cm}\PBOXa\\
&& + ~{\rm pinchings}
\end{eqnarray*}
which again indicates that the ${\cal O}(\epsilon)$ part of $I_1$ is necessary
to determine the ${\cal O}(\epsilon^0)$ part of $I_2$.
In fact this additional part is not very difficult to obtain either by 
considering the differential equations for $I_1$ and $I_2$ at $t=-s$ \cite{gehrmann}
or by explicit evaluation of the Mellin-Barnes integrals \cite{tausk}.
As expected, the $\epsilon$ expansion for $I_3$ through to ${\cal
O}(\epsilon^0)$ contains quadrilogarithms at worst. 

\section{Results}
Using the analytic expansions around $D\sim 4$ we can evaluate the 
leading colour two-loop amplitudes for $q \bar q \to q^\prime \bar q^\prime$.  
Expanding both the integrals and coefficients $a_1$ -- $a_{14}$ 
we find that the
leading singularities are proportional to tree level,
\begin{eqnarray*}
\lefteqn{{\cal M}_2 {\cal M}_0^\dagger + {\cal M}_2^\dagger {\cal M}_0
=}\\
&&
\frac{4N^2}{(4\pi)^D}
\frac{\Gamma(1+\epsilon)^2\Gamma(1-\epsilon)^4}{\Gamma(1-2\epsilon)^2}
\frac{1}{\epsilon^4}\left(\frac{-t}{\mu^2}\right)^{-2\epsilon}
~|{\cal M}_0|^2 \\
&& + {\cal O}\left(\frac{1}{\epsilon^3}\right).
\end{eqnarray*}
The leading pole has the same coefficient as the square of the 
one-loop amplitude.  The remaining pole and finite contributions are in the
process of being checked.

\section{Summary}
We have made a study of two-loop amplitudes of massless $2 \to 2$ scattering by
considering $q \bar q \to q^\prime \bar q^\prime$ as a trial process.  At
leading colour, only planar graphs contribute and we have expressed the
amplitude as a sum over the basis set of two-loop master integrals.  It turns
out that for this process,    in reducing the tensor integrals to scalars,
factors of  $1/(D-4)$ are generated multiplying both of the Smirnov-Veretin
planar box graphs, $I_1$ and $I_2$.  These factors do not cancel in the
physical process thereby necessitating a knowledge of both $I_1$ and $I_2$ to
${\cal O}(\epsilon)$. Alternatively, if one uses the basis $I_1$ and the
irreducible numerator integral $I_3$, the factors of $1/(D-4)$ disappear.    Of
course, evaluating $I_3$ also requires the  ${\cal O}(\epsilon)$ of $I_1$ but
this has now been calculated \cite{gehrmann,tausk}. We therefore have the
ingredients to evaluate the leading colour  two-loop amplitude for $q \bar q
\to q^\prime \bar q^\prime$ for $D \sim 4$  and the leading singularities agree
with expectations.

\subsection*{Acknowledgements}
We thank C. Anastasiou, D. Broadhurst, T. Gehrmann, C. Oleari and J.B. Tausk
for discussions and useful suggestions. 
M.E.T. acknowledges financial support
from CONACyT and the CVCP. We gratefully acknowledge the support of
the British Council and German Academic Exchange Service under ARC
project 1050.

\end{document}